\documentclass[12pt]{article}
\usepackage{amssymb,graphicx}

\def\p{\mathrm{p}}

\def\b{\mathrm{b}}
\def\d{\mathrm{d}}
\def\v{\mathrm{v}}
\def\m{\mathrm{m}}
\def\mp{{\mathrm{m}^\prime}}

\def\vr{\vec{r}}
\def\vn{\vec{n}}

\def\vf{\vec{F}}
\def\va{\vec{A}}

\def\vp{\vec{p}}

\def\cf{{\cal F}}

\begin{document}

\begin{flushleft}
  {\footnotesize \bf Gakuto International Series}\\
  {\footnotesize \bf Mathematical Sciences and Applications (2011)}\\
  {\footnotesize in press} 
\end{flushleft}
\vspace{1.5cm}

                   \begin{center}
   {\large \bf THE STUDY OF MACROSCOPIC DYNAMICS OF NANO PROCESSES IN BALL MILLS THROUGH NANO-SCALE SIMULATION}\\[1cm]
                   \end{center}

                   \begin{center}
                  {\sc G. K. Sunnardianto$^a$, L.T. Handoko$^{a,b}$}\\
 (gagus@teori.fisika.lipi.go.id, handoko@teori.fisika.lipi.go.id, 
handoko@fisika.ui.ac.id)
\vspace{0.2cm}

$^{a)}$Group for Theoretical and Computational Physics, Research Center for
Physics, Indonesian Institute of
Sciences\footnote{http://teori.fisika.lipi.go.id}, Kompleks Puspiptek Serpong,
Tangerang 15310, Indonesia.\\
$^{b)}$Department of Physics, University of 
Indonesia\footnote{http://www.fisika.ui.ac.id}, Kampus UI Depok, Depok 16424,
Indonesia.\\
                    \end{center}
\vspace{2cm}
         
\noindent
{\bf Abstract.} 
Numerical simulation for comminution processes inside the vial of ball
mills are performed using Monte Carlo method. The internal dynamics is
represented by recently developed model based on hamiltonian involving the
impact and surrounding electromagnetic potentials. The paper is focused on
investigating the behaviors of normalized macroscopic pressure, $P/{P_0}$, in
term of system temperature and the milled powder mass. The results provide
theoretical justification that high efficiency is expected at low system
temperature region. It is argued that keeping the system temperature as low as
possible is crucial to prevent agglomeration which is a severe obstacle for
further comminution processes.
         
\vfill
\noindent
------------------------------------------------------------------------------
\\
{\footnotesize Received xxxxxxxxxx, 2011.\\
This work is supported by Riset Kompetitif LIPI FY 2010.\\
AMS Subject Classification 70-08,70F99}

\newpage

\section{Introduction}

Ball mills have been deployed as a simple top-down approach for comminution
processes up to nanometer scale. Although its practicality, ball mills involve
many parameters related to both external and internal dynamics of the
equipments. All of them unfortunately lead to severe uncertainties and become
some major obstacles to perform an efficient milling.

Meanwhile, more theoretical approaches based on mathematical description are
expected to overcome such problems. Many efforts have been done to develop 
modelization of ball mill equipments 
\cite{mishra92,mishra94,mishra95,mishra01,poschel}. Through the modeling
approach and its subsequent simulation, one expects to be able to obtain prior
information and constraints to optimize experimental strategy.

In general, the experimental works using ball mill equipments face the following
problems :
\begin{itemize}
\item Choosing an appropriate parameter set related to the milled materials for
particular case of experiments. It could be the ball size, the initial size of
milled powders, the number of ball, the initial quantity of powders, material
characteristics of ball and also powders, and so forth.
\item The milling time for particular characteristics and quantities of balls
and powders.
\item The design and geometrical motions of vial itself. Although there are 
various types of ball mills, most of them have not been developed based on prior
comprehensive simulations.
\item On the other hand, the experimental measurements on the internal dynamics
of vial are almost impossible.
\end{itemize}
The question is then how to overcome these problems ? Is there any smart
solution for these obstacles ?

In our recent works, a novel model to describe the internal dynamics and to
relate it with surrounding environment has been proposed
\cite{handoko,handoko2}. In contrast to the semi-empirical approaches, the
model does not require prior experimental data or simulation results to fit the
parameters \cite{manai,davis,maurice1990,maurice1996}, nor huge computational
power as in some more empirical approaches \cite{delogu,wang,concas}. Actually
our model combines the deterministic approach for milling bodies motion, and the
statistical approach to relate them with external macroscopic physical
observables, in particular system temperature.

In this paper, however the focus is put on investigating the behaviors of normalized macroscopic pressure, $P/{P_0}$. The observable is important and accessible in most of real experiments. Particular interest is investigating its dependencies on the system temperature and the evolution milled powder mass as well.

The paper is organized as follows. First, after this introduction the model is 
briefly explained. Before summarizing the results, numerical analysis and
simulation for the normalized pressure in term of system temperature are
discussed.

\section{Model and simulation}

Rather solving a set of equation of motions (EOMs) governing the whole dynamics
as always done in conventional approaches, the dynamics is described in a
hamiltonian involving all considerable potentials in the system and its
surrounding environment. 

In the model, the dynamics of each 'matter' in the system, i.e. balls and
powders inside the vial, are described by a hamiltonian $H_\m(\vr,t)$. The index
$\m$ denotes the powder ($\p$) or ball ($\b$) and $\vr = (x,y,z)$. The
hamiltonian contains some terms representing all relevant interactions working
on the matters inside the system as follow \cite{handoko2}, 
\begin{equation}
 H_\m = H_0 + V_{\m-\m} + V_{\m-\v} + V_{\m-\m^\prime} + V_\mathrm{ext} \; ,
\label{eq:h}
\end{equation}
with $\v$ denotes the vial, while $H_0$ is the free matter hamiltonian
containing the kinetic term,
\begin{equation}
 H_0 = \frac{1}{2 m_\m} \sum_{i=1}^{n_\m} \left| \left( \vp_\m \right)_i
\right|^2 \; ,
\label{eq:h0}
\end{equation}
where  $n_\m$ is the matter number, $m_\m$ and $\vp_\m$ are the matter mass and
momentum respectively. Throughout the paper we assume that the mass or size
evolution of matters is uniform for the same matters.

The matter self-interaction $V_{\m-\m}$, the matter--vial interaction
$V_{\m-\v}$ and the interactions between different matters may be induced by,
for instance,  the impact ($V^{\mathrm{imp}}$) potential,
\begin{eqnarray}
 V^{\mathrm{imp}}_{\m-\mp}(\vr,t) & = & -\sum_{i=1}^{n_\m} \sum_{j=1}^{n_\mp} 
\int_0^{\left(\xi_{\m\mp}\right)_{ij}} \d \left(\xi_{\m\mp}\right)_{ij} \, \vn
\cdot \left( \vf^{\mathrm{imp}}_{\m\mp} \right)_{ij} \; , 
 \label{eq:vimp}
\end{eqnarray}
with $\m,\mp : \v, \p, \b$ and $\vn$ is the unit normal vector. The potential 
should in fact represent the whole classical dynamics among the matters, i.e.
the impact forces among balls and powders. Here the impact force is 
dominated by its normal component \cite{concas},
\begin{equation}
 \vf^{\mathrm{imp}}_{\m\mp}(\vr,t) = \left [ \frac{2 \Upsilon_{\m\mp}}{3 (1 -
v_{\m\mp}^2)} \sqrt{R_{\m\mp}^\mathrm{eff}} \left ( \xi_{\m\mp}^{{3}/{2}} +
\frac{3}{2} A_{\m\mp} \sqrt{\xi_{\m\mp}} \, \frac{\d \xi_{\m\mp}}{\d t}\right)
\right] \vn \; .
\label{eq:fn}
\end{equation}
Here, $\Upsilon_{\m\mp}$ is the Young
modulus, $v_{\m\mp}$ represents the Poisson ratio of the sphere material,
$R_{\m\mp}^\mathrm{eff} = {(R_\m R_\mp)}/{(R_\m + R_\mp)}$ is the effective
radius, while $\xi_{\m\mp} = R_\m + R_\mp - | \vr_\m - \vr_\mp|$ is the
displacement with $R_\m$ is the radius of interacting matter. $A$ is a
dissipative parameter given in \cite{brilliantov,landau,hertzsch} containing the
viscous constant $\eta_\m$.

In fact, there are another potentials like Coulomb and gravitational potentials
which may influence on the system. However, those contributions should be
considerably tiny due to its neutral charges and tiny masses as well \cite{handoko2}. 

Incorporating the effect of external electromagnetic field surrounding the
system for charged matters shifts the kinetic term in Eq. \ref{eq:h0} to be
\cite{handoko2}, 
\begin{equation}
 H_0 \longrightarrow H_{0 + \mathrm{EM}} = \frac{1}{2 m_\m} \sum_{i=1}^{n_\m}
\left| \left( \vp_\m \right)_i - Q_\m \, \va \right|^2 + n_\m \, Q_\m \, \phi \;
,
\label{eq:h0em}
\end{equation}
with electromagnetic (scalar and vector) potential $(\phi,\va)$ and the matter
charge $Q_\m$.

From now, let us focus only on the dynamics of powders which is our main
interest in the sense of comminution process. From Eqs. (\ref{eq:h}),
(\ref{eq:h0}) and (\ref{eq:vimp}), the total
hamiltonian for the powder in the model is,
\begin{eqnarray}
 H_\p & = & \frac{1}{2 m_\p} \sum_{i=1}^{n_\p} \left| \left( \vp_\p \right)_i -
Q_\p \, \va \right|^2 + n_\p \,Q_\p \, \phi  
	\nonumber \\
	&& - \frac{1}{2} \sum_{i(\neq j)=1}^{n_\p} \sum_{j=1}^{n_\p} 
\int_0^{\left(\xi_{\p\p}\right)_{ij}} \d \left(\xi_{\p\p}\right)_{ij} \, \vn
\cdot \left( \vf^{\mathrm{imp}}_{\p\p} \right)_{ij} 
	\nonumber \\
	&& - \sum_{m:\b,\v}\sum_{i=1}^{n_\p} \sum_{j=1}^{n_\m} 
\int_0^{\left(\xi_{\p\m}\right)_{ij}} \d \left(\xi_{\p\m}\right)_{ij} \, \vn
\cdot \left( \vf^{\mathrm{imp}}_{\p\m} \right)_{ij} \; ,
\label{eq:hp}
\end{eqnarray}
for $Q_\p \neq 0$. The last two potentials represent the total impact potential
among powders; powders and vial; powders and balls respectively. Obviously we do
not need to take into account the ball self-interaction
$V^{\mathrm{imp}}_{\b-\b}$ nor ball-vial interaction $V^{\mathrm{imp}}_{\b-\v}$.
This is actually the advantage of using hamiltonian method.

Relating a hamiltonian with macroscopic physical observables can be realized
through the partition function known in statistical mechanics, 
\begin{equation}
 Z_\m = \int \prod_{i=1}^{n_\m} \d \vp_i \, \d \vr_i \; \mathrm{exp} \left[ 
 -\int_0^\beta \d t \, H_\m
 \right] \; ,
\label{eq:z}
\end{equation}
for a canonical ensemble of matter m governed by a particular hamiltonian
$H_\m$. Here, $\beta \equiv 1/{(k_B T)}$
with $k_B$ and $T$ are the Boltzman constant and absolute temperature. 
Further one can calculate, for instance the normalized pressure as, 
\begin{equation}
 P'_\m = \frac{\ln Z_\m}{\ln {Z_0}_\m}
\; .
\label{eq:nf}
\end{equation}

\begin{figure}[t]
        \centering 
	\includegraphics[width=\textwidth]{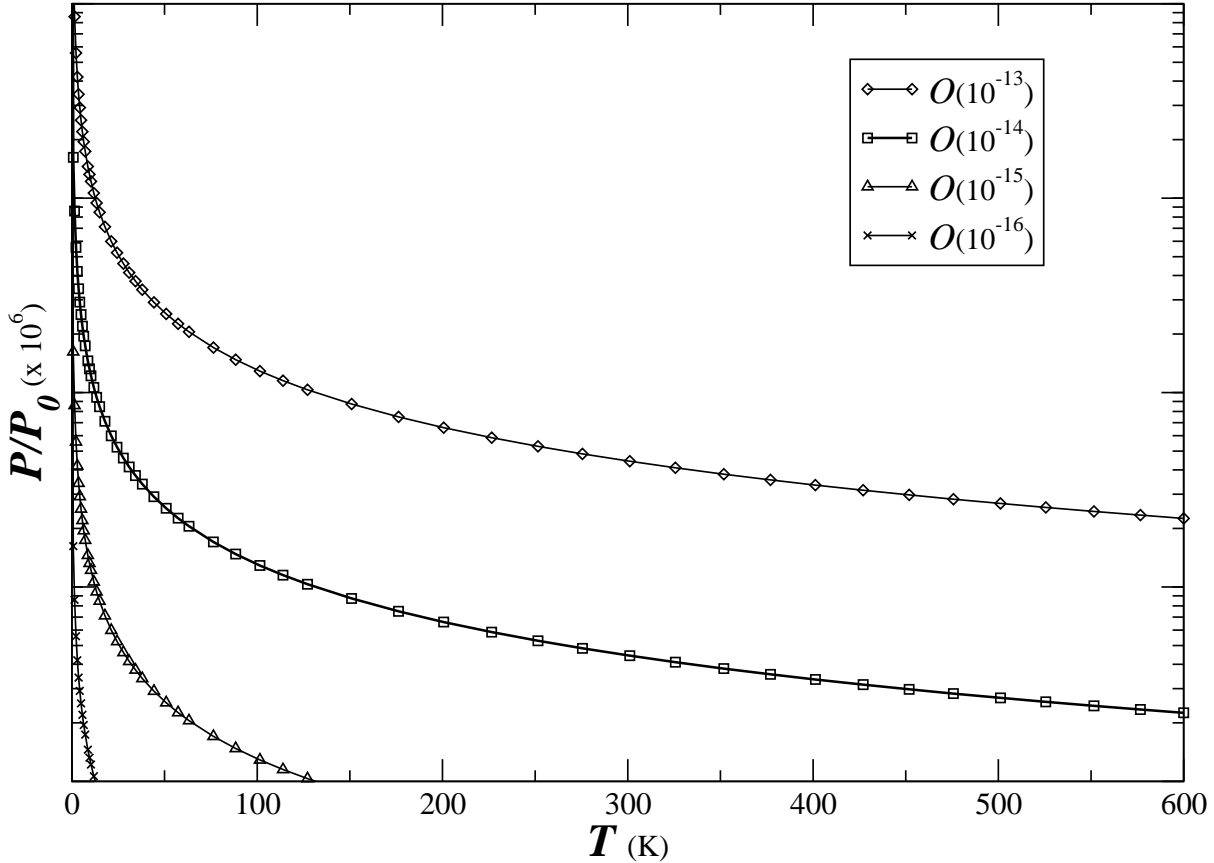}
        \caption{The normalized macroscopic pressure, $P/{P_0}$, as a function
of temperature for typical order values of $\cf = O(10^{-13}) \sim
O(10^{-16})$.}
        \label{fig:pt}
\end{figure}

\begin{figure}[t]
        \centering 
	\includegraphics[width=\textwidth]{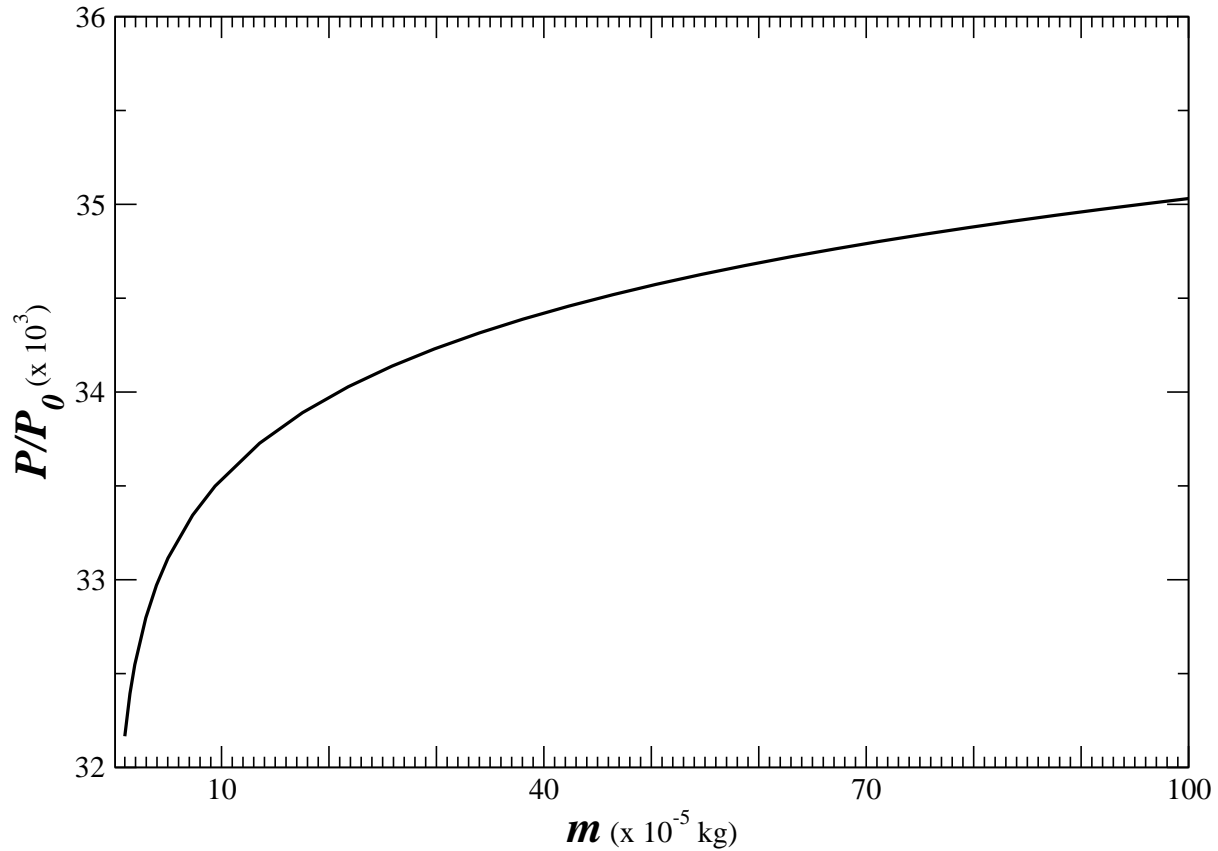}
        \caption{The normalized macroscopic pressure, $P/{P_0}$, as a function
of powder mass for $\cf \sim O(10^{-14})$ at typical room temperature.}
        \label{fig:pm}
\end{figure}

Integrating out the time component at finite temperature, one obtains
\cite{handoko2}, 
\begin{equation}
  P'_\p = 1 - \beta \, \cf \,  \ln^{-1} \left( \frac{2 \, m_\p \pi}{\beta}
\right) \; ,
\label{eq:pf}
\end{equation}
respectively with,
\begin{eqnarray}
 \cf & \equiv & 2 \,  \int \prod_{i=1}^{n_\p} \d \vr_i  \left[ 
  Q_\p \, \phi  
 - \frac{2}{15 \, n_\p} \sum_{i(\neq j)=1}^{n_\p} \sum_{j=1}^{n_\p}
\frac{\Upsilon_{\p\p}}{1 - v_{\p\p}^2} \sqrt{R_{\p\p}^\mathrm{eff}} \,
\left(\xi_{\p\p}\right)_{ij}^{{5}/{2}} 
  \right. \nonumber \\
  && \left. - \frac{4}{15 \, n_\p} \sum_{\m:\b,\v} \sum_{i=1}^{n_\p}
\sum_{j=1}^{n_\m} \frac{\Upsilon_{\p\m}}{1 - v_{\p\m}^2}
\sqrt{R_{\p\m}^\mathrm{eff}} \, \left(\xi_{\p\m}\right)_{ij}^{{5}/{2}} 
  \right] \; .
\label{eq:faux}
\end{eqnarray}
Eq. (\ref{eq:pf}) provides a general behavior for temperature-dependent pressure
in the model, while the geometrical structure and motion of vial is absorbed in
the function $\cf$. 

This is the final result which is ready to be evaluated further using numerical
approaches like Monte Carlo. 

\section{Results and summary}

In this paper, the integral in Eq. (\ref{eq:faux}) is performed using Monte
Carlo technique. Simulation on the system temperature dependencies of
normalized pressure is shown in Fig. \ref{fig:pt}. Secondly its dependencies on
matter mass is given  in Fig. \ref{fig:pm}.

The simulation is done for steel material in a typical geometry of spex mixer /
mill with resolutions on 3-dimensional space $\sim 10^4 \times 10^4 \times
10^4$, vial length $l = 50$ mm, vial radius $r_\v = 10$ mm, shaft-arm length $L
= 200$ mm and ball radius $R_\b = 5$ mm.

From the figures one can conclude that high efficiency is expected at low system
temperature region as can be understood from natural sense since the higher is
temperature, the agglomeration phenomena occurs which prevents further
comminution processes. In another words, the higher is temperature, the pressure
is getting down which lead to lower impact energy. Therefore these results
suggest that in general one should keep optimized working temperature as long as
possible to keep high energy milling. 

\section*{Acknowledgments}

The authors greatly appreciate inspiring discussion with N.T. Rochman and A.S.
Wismogroho throughout the work. This work is funded by the Riset Kompetitif LIPI
in fiscal year 2010 under Contract no.  11.04/SK/KPPI/II/2010. LTH would like
to thank the Organizer of ISCS 2011 at Kanazawa University for financial
support and warm hospitality.

\bibliographystyle{unsrt}
\bibliography{ISCS2011}

\end{document}